\documentclass[twoside]{article}

\usepackage{PRIMEarxiv}

\usepackage[utf8]{inputenc} 
\usepackage[T1]{fontenc}    
\usepackage{hyperref}       
\usepackage{url}            
\usepackage{booktabs}       
\usepackage{amsfonts}       
\usepackage{nicefrac}       
\usepackage{microtype}      
\usepackage{lipsum}
\usepackage{fancyhdr}       
\usepackage{graphicx}       
\usepackage{caption}
\usepackage{multirow}
\usepackage{wrapfig}
\usepackage{float}
\usepackage{blindtext}
\graphicspath{{media/}}     

\title{Mining Tourism Experience on Twitter: A case study
}

\author{
  Davide Stirparo, Beatrice Penna, Mohammad Kazemi, Ariona Shashaj \\
  Network Contacts \\
  Rome, Italy \\
  \texttt{\{davide.stirparo, beatrice.penna, mohammad.kazemi, ariona.shashaj\}@network\-contacts.it} \\
}

\begin{document}
\maketitle

\begin{abstract}
With the increase of digital data and social network platforms the impact of social media science in driving company decision related to product/service features and customer care operations is becoming more crucial. In particular, platform such as Twitter where people can share experience about almost everything can drastically impact the reputation and offering of a company as well as of a place or tourism site. Text mining tools are researched and proposed in literature in order to gain value and perform trend topics and sentiment analysis on Twitter. As data are the fuels for these models, the "right" ones, i.e the domain-related ones makes a difference on their accuracy.
In this paper, we describe a pipeline of \textit{DataOps / MLOps} operations performed over a tourism related Twitter dataset in order to comprehend tourism motivation and interest. The gained knowledge can be exploit, by the travel/hospitality industry in order to develop data-driven strategic service, and by travelers which can consume relevant information about tourist destination.
\end{abstract}


\section{Introduction}
\label{intro}
The \textit{Customer Experience} domain covers many aspects of a company offering such as, customer care quality, advertising, as well as products and services features~\cite{meyer2007understanding}. In this context, understanding what people think and feel about products/services is crucial in order to deliver qualitative customer care operations, target oriented advertising and improved products/services features.
On the other hand, social medias where people share information and experience about products, services, places and events are the virtual mining field to gain value and achieve data-driven customer experience process.\\
Twitter, an online social networking service which allow people to share short messages regarding topics of interest, named "tweet", with its 397 million of active users \footnote{https://www.statista.com/statistics/272014/global-social-networks-ranked-by-number-of-users/} is on the top ten of most used social media. The Twitter REST API and Streaming API \footnote{https://dev.twitter.com/docs/api/1.1} provide static data such as user demographic information, as well as streaming data i.e "tweets". \\
In social science, there are many efforts focused on gaining value from these data, mostly related to topics and sentiment analysis~\cite{kumar2014twitter}~\cite{antonakaki2021survey}. Research works such as \cite{khan2021twitter},~\cite{hassonah2020efficient} are focused on developing and improving techniques related to sentiment classification and trend topic mining, whereas works like~\cite{yigitcanlar2021smart},~\cite{garcia2021topic} (and many others) reports applied user cases of sentiment analysis and topic detection on Twitter's data.\\
The general workflow which characterized Twitter and in general text mining analytics includes four major phases: data collection /ingestion (operations related to the data retrieval process), pre-processing (cleaning and filtering process), analysing (data mining operations) and serve the results (visualizing tools /operations). This work is focused on the pre-processing and analysing phase. \\
%
In this paper, we describe a text-mining analytic pipeline designed to mine knowledge from tourist content posted on Twitter.
Our study is driven by the final goal to understand tourist motivations in travel and hospitality industry in Puglia\footnote{Southern region in Italy - https://en.wikipedia.org/wiki/Apulia}.
The designed pipeline include:
\begin{itemize}
    \item A keyword-base data injection phase.
    \item Pre-processing phase base on NLP(\textit{Natural Language Processing}) technique to obtain a clean-domain related contents.
    \item Data mining analytic steps defined to gain knowledge from collected tweets.
\end{itemize}
%
The rest of the paper is organized as follows. In Section~\ref{sec:soa} we review related works and motivate our proposal. In~\ref{sec:data_cleaning} processes involved in the data preparation phase are described. In Section~\ref{sec:Social network analysis } we show the designed analytical steps. Finally, conclusions remarks and future works are given in Section~\ref{sec:conclusion}.

\section{Motivations and related works}
\label{sec:soa}
In literature, there are different works focused on Twitter and social media analysis.
In~\cite{singh2016role}, with the final aim to bind slang words to vocabulary's ones colored with a sentiment polarity, the authors proposed a methodology for text normalization based on NLP operations which are shown to increase the accuracy of sentiment classification. \\
In~\cite{jianqiang2017comparison}, ~\cite{symeonidis2018comparative}, ~\cite{naseem2020survey} comparative studies of different text pre-processing techniques have been performed which show how repeated letters and acronyms found in tweets impact the accuracy of tweet sentiment.\\
An exploratory study has been conducted in~\cite{reyes2018understanding} in order to identify social, economical, and other factors that mostly concern Twitter users about environmental and public health. The study describes a \textit{Social Network Analysis} (SNA) approach based on Textual and Sentiment Analysis.
In the last two years, considering the pandemic emergency of Covid-19, many works on Twitter analysis have been focused on tracking topic trends and sentiment feelings. \\
In order to consider a domain specific dataset, the authors in~\cite{boon2020public} have performed topic modeling, keywords extraction, and sentiment analysis over a Twitter dataset Covid-19 related collected using specific hashtags (\textit{\#covid\_19} and similar).\\
In~\cite{ordun2020exploratory} authors performed an exploratory analysis of a topic and network dynamics over a Twitter dataset related to Covid-19. The dataset was collected considering medical related keywords. \\
A modular framework for dynamic topic modeling and sentiment classification has been described in~\cite{yin2020detecting} and results are shown considering a dataset collected based on Covid-19 keywords.\\
This works have in common the domain-related dataset collected through specific keywords \ hashtags like those related to the coronavirus. \\ 
The deep comprehension of tourism need can be exploited by the travel / hospitality as well as by the regional administration to further improve offered service and the overall tourist's satisfaction. 
Considering that our main focus is to develop an analytic framework related to regional tourism, the target dataset has been collected through the free streaming API, from November 2019 to April 2021, by harvest tweets through keywords related to a specific region in Italy, Puglia.
\\
One of the first challenge the collected dataset posed is that, even though the harvest keywords were selected considering particular features of the target region, the contents of the tweets can be generic or in some cases not related at all even with Italy. 
On the other hand, restricting the collection to just a bunch of tourism-related hashtags results in a poor dataset. 
Considering this, the initial data pre-processing phase described in this work has a two-fold objective, reducing a generic Twitter dataset to a domain-related one and augmenting the domain-related dataset with content that can be excluded from the hashtag base collecting process.\\
Further, the designed analytics steps applied on the domain-related dataset has been designed with the final aim to understand tourist motivations and interests. 

\section{Data Preparation}
\label{sec:data_cleaning}
Tweets contain contrasting and different content and often they carry hashtags and emojis which all make them very complicated to work with. Consequently, finding a specific topic and gaining adequate data among all them, require a proper data cleaning method that is able to deal with their complexities. \\
The first challenge of the work that we are presenting here was to find the best approach to clean our dataset, in order to obtain only tweets about tourism in Puglia. The proposed methodology it is based on an enhanced topic modeling approach executed in order to identify and restrict the target dataset to tourist related topics. 

\subsection{Data Processing}
By using Twitter-API, we collected around 900.000 tweets, from which we only selected texts in English and Italian (around 190.000 tweets in English and around 540.000 in Italian). 
In the first step we removed duplicate tweets as well as special characters and stopwords, then we did the tokenization and the lemmatization. Furthermore, the process of extracting useful information from the given words by using machine learning requires the string/text to be converted into a set of real numbers (a vector). The process of converting text documents into scalar vectors is called \textit{Vectorization}. Here we used the TF-IDF(\textit{Term-Frequency Inverse Document Frequency}) approach .

\subsection{Topic Modeling}
Latent Dirichlet Allocation (LDA)~\cite{blei2003latent} is one of the most popular and most used topic modeling technique. Nevertheless,  there is some uncertainty about the validity and reliability of the LDA results, and it needs to be properly tuned in order to find the best values for hyperparameters to get the optimum result~\cite{shashaj2021cea}.\\
The table~\ref{tab:LDA English and Italian} shows the output after applying the LDA on the whole dataset. Even though the model that we used was a tuned one, the results were ambiguous and it was not possible to find any clear topic from it. In order to improve the quality, we propose here a new approach that helped us to filter only the tweets  we were interested in.
\vspace{-5pt}
 \begin{table}[ht]
 	\centering
 	\scalebox{0.6}{
 		\begin{tabular}{|p{1.6cm}|p{7.5cm}|p{7.5cm}|}
 			\hline
 			\textbf{Topic} & \textbf{English keywords } & \textbf{Italian keywords}\\
 			\hline
 			\hline
 			  Topic 0 & gargano, match, wwenxt, nxt, ciampa, johnny, vs, takeover, cole, balor & puglia, bari, quoti, rep, lagazzettaweb, burrata, ilmattinofoggia, corrmezzogiorno, fattoquotidiano, sud \\
 			\hline
 			  Topic 1 & puglia, italy, salento, apulia, region, italian, travel, wine, beautiful &  puglia, campania, sicilia, calabria, zona, regioni, toscana, lombardia, veneto, sardegna \\
 			\hline
 			 Topic  2 & year, otranto, puglia, castle, go, day, moment, say, amp, read, back &  puglia, salento, gargano, taranto, otranto, weareinpugliapuglia, news, post, madeintaranto, complimenti \\
 			\hline
 			 Topic  3 & gargano, julia, americanidol, taranta, save, love, top, vote, trulli, heart & puglia, salento, mare, italy, italia, foto, trulli, gargano, gt, buongiorno \\
 			\hline
 			 Topic  4 & gargano, like, johnny, get, good, ciampa, heel, think, go, make & puglia, casa, solo, persone, due, bari, andare, fare, giorni, milano \\
 			\hline
 			  Topic 5 & gargano, johnny, go, get, win, johnnygargano, see, gt, want, lose & salento, notizie, puglia, lecce, san, gargano, ansa, foggia, bari, gallipoli \\
 			\hline
 		     Topic 6 & burrata, cheese, tomato, amp, salad, make, fresh, pizza, eat, basil &  puglia, emiliano, pd, regionali, regione, fitto, presidente, governo, candidato, elezioni \\
 			\hline
 		    Topic 7 & vs, gargano, johnny, lee, team, priest, keith, trulli, nxt, american & puglia, coronavirus, covid, casi, regione, nuovi, positivi, bari, contagi, oggi \\
 		    \hline
 		    Topic 8 & gargano, vs, ciampa, cole, match, johnny, nxt, well, adam, wwe & puglia, andria, salento, solo, fare, anno, anni, burrata, dire, bene, ora  \\
 		    \hline
 		    Topic 9 & puglia, tree, apulia, olive, italy, libra, amp, christmas, bari, like & puglia, salento, matera, bari, basilicata, taranto, brindisi, primitivo, lecce, foggia  \\
 		    \hline
 		\end{tabular}}
 	\caption{LDA English and Italian}
 	\label{tab:LDA English and Italian}
 \end{table}

\subsection{Propose Method}
\label{data_expl}
\begin{figure}[h]
\vspace{-20pt}
 \begin{center}
    \includegraphics[width=9cm, height=10cm]{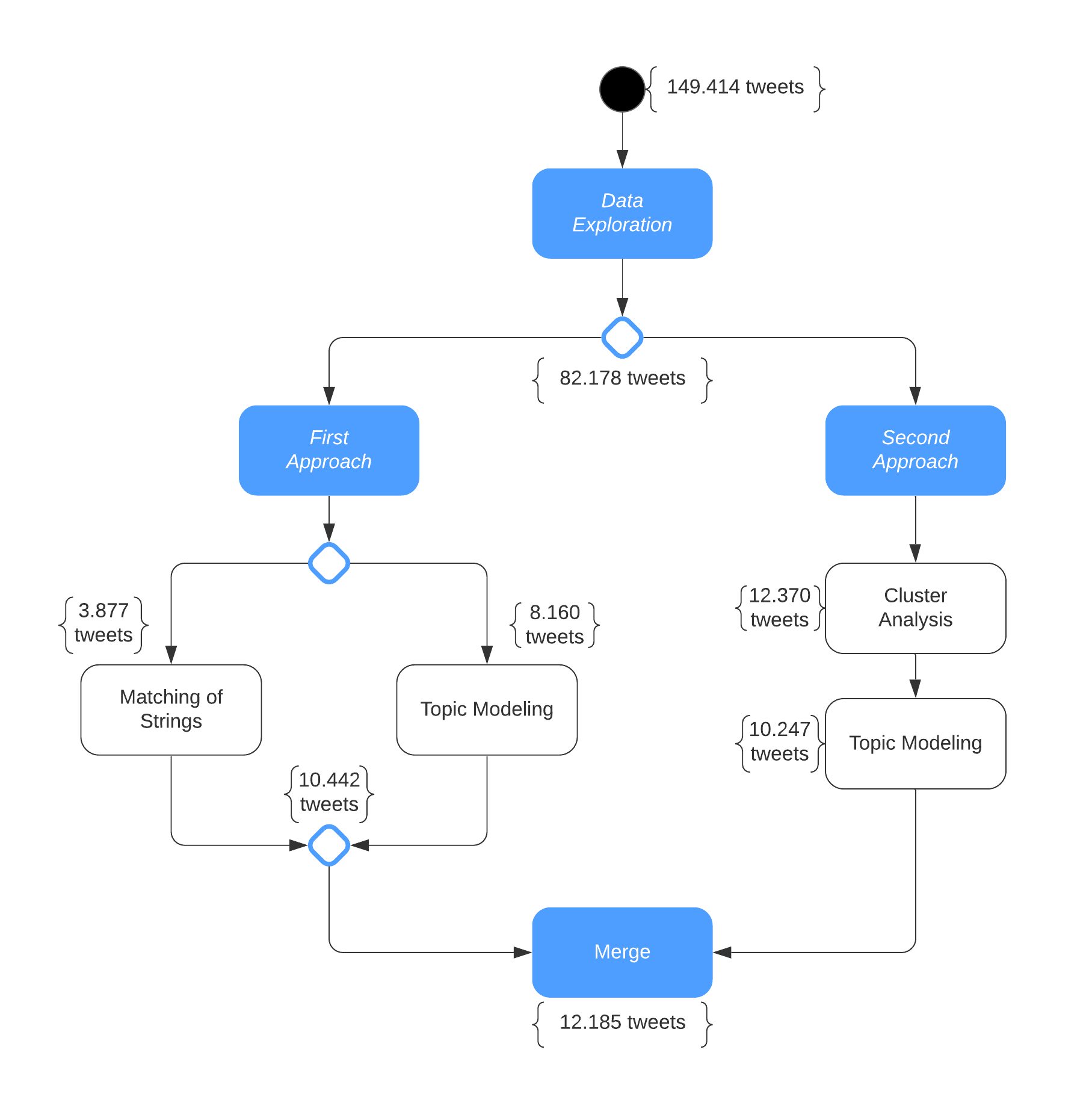}
 \end{center}
 \caption{Enhanced topic-modeling workflow. }
 \label{fig:Workflow}
 \vspace{-10pt}
\end{figure}

In figure~\ref{fig:Workflow} we show the proposed topic modeling enhanced workflow designed to improve topic modeling results and identify topic related to tourism. The amount of tweets shown refers to the English dataset but we followed the same steps for the Italian as well.
\textbf{Data Exploration:} In the data exploration, by using tokens resulted from the NLP pre-processing phase, we gathered the top 1.000 most frequent words and hashtags from the dataset. We labeled them manually and placed each token and its frequency into two dictionaries, \textit{Tourism} and \textit{Not Tourism}. Subsequently, by using these dictionaries, we are going to define a classification method in order to clustered tweets as \textbf{Tourism} related (table~\ref{tab:sample-table}) and not (\textbf{Not Tourism}). 
\\
 \subsubsection{First Approach: Matching Strings}
       
\textbf{Matching of Strings step:} By using the selected words in the data exploration, we kept only tweets that contained at least 3 occurrences of terms belonging to the \textbf{Tourism} words. It allowed us to select only small fraction of touristic tweets.\\ 

\textbf{Topic Modeling :} We applied LDA to label each tweet with the argument covered by it. In detail, we applied LDA several times by repeating it on the texts that belonged to the tourism topic. In this way we were able to clean our data as much as possible. For instance for the English dataset at the first time we obtained 3 topics which only one of them belonged to tourism and it consisted of 27.000 tweets. \\
Then we repeated the LDA again on those tweets, and it gave us 8.160 tweets.
\begin{table}[ht]
	\centering
	\scalebox{0.7}{
		\begin{tabular}{|c|c|c|c|}
			\hline
			\multicolumn{2}{|c|}{Italian} & \multicolumn{2}{|c|}{English}\\
			\hline
			\textbf{Tourism} & \textbf{Not Tourism} & \textbf{Tourism} & \textbf{Not Tourism}\\
			\hline
			\hline
            lungomare & fattoquotidiano & enjoy & nxtciampa\\
            prenota    &  mascherina & entertain & nxtgab\\
            resort &  giorgiameloni & holiday & nxttakeover\\
            italianholidays & tgrai & trip & wrestlemania\\
            lacasadegliartisti & quarantena & travel & wrestlefeatures\\
            apuliatravels & iorestoacasa & beach & horace\\
            borghiitalia & droga & placevisit & walpole\\
			\hline
		\end{tabular}}
	\caption{Words sample from each category}
	\label{tab:sample-table}
\end{table}
Finally, we merged the outputs of \textit{Matching of Strings} and \textit{Topic Modeling} and we removed tweets that were present in both, so for example we ended up with 10.422 tweets in English dataset.

\subsubsection{Second Approach:Clustering}

Here we used k-means clustering in order to identify cluster of tweets that belonged to tourism. 
By maximizing the \textit{silhouette score} we achieved several clusters. The most frequent words for each cluster helped us to find tweets related to tourism. Moreover, to have more cleaned tweets, we applied here the LDA model as described previously.\\
\begin{figure}[ht]
  \centering
  \begin{minipage}[b]{0.48\textwidth}
     \includegraphics[width=1\linewidth]{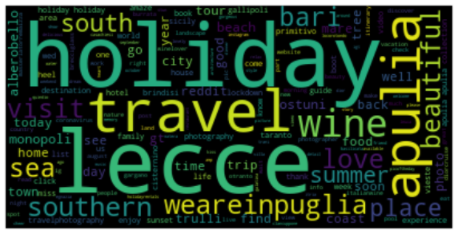}
     \caption{WordCloud Tourism English}
     \label{Fig:tourism en}
   \end{minipage}
   \hfill
   \begin{minipage}[b]{0.48\textwidth}
     \includegraphics[width=1\linewidth]{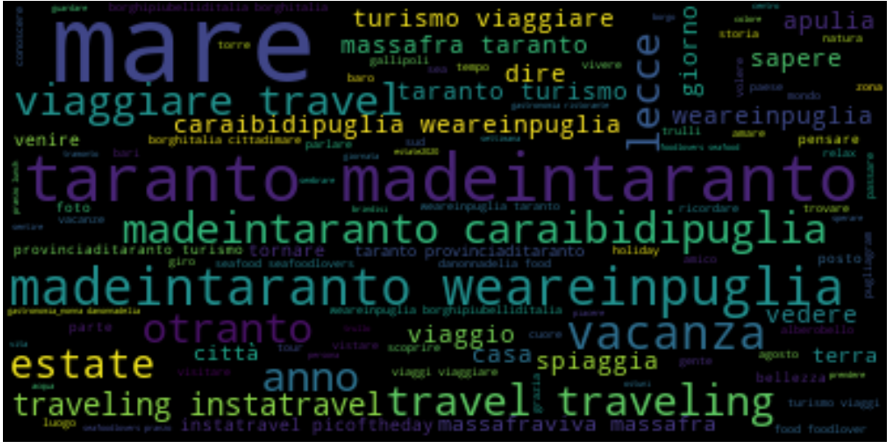}
     \caption{WordCloud Tourism Italian}
     \label{Fig:tourism it}
   \end{minipage}
\end{figure}


After the execution of the first and second approaches, we obtained two datasets that are merged into one. By removing the duplicates, we obtained 12.185 touristic tweets for English (figure ~\ref{Fig:tourism en}) which was only 8.6\% of the entire dataset, and 32.505 which is 8.8\% for Italian dataset (figure ~\ref{Fig:tourism it}).

\subsection{Sub categories of tourism}

\begin{figure}[h]
  \vspace{-20pt}
  \begin{center}
    \includegraphics[width=10cm, height=6cm]{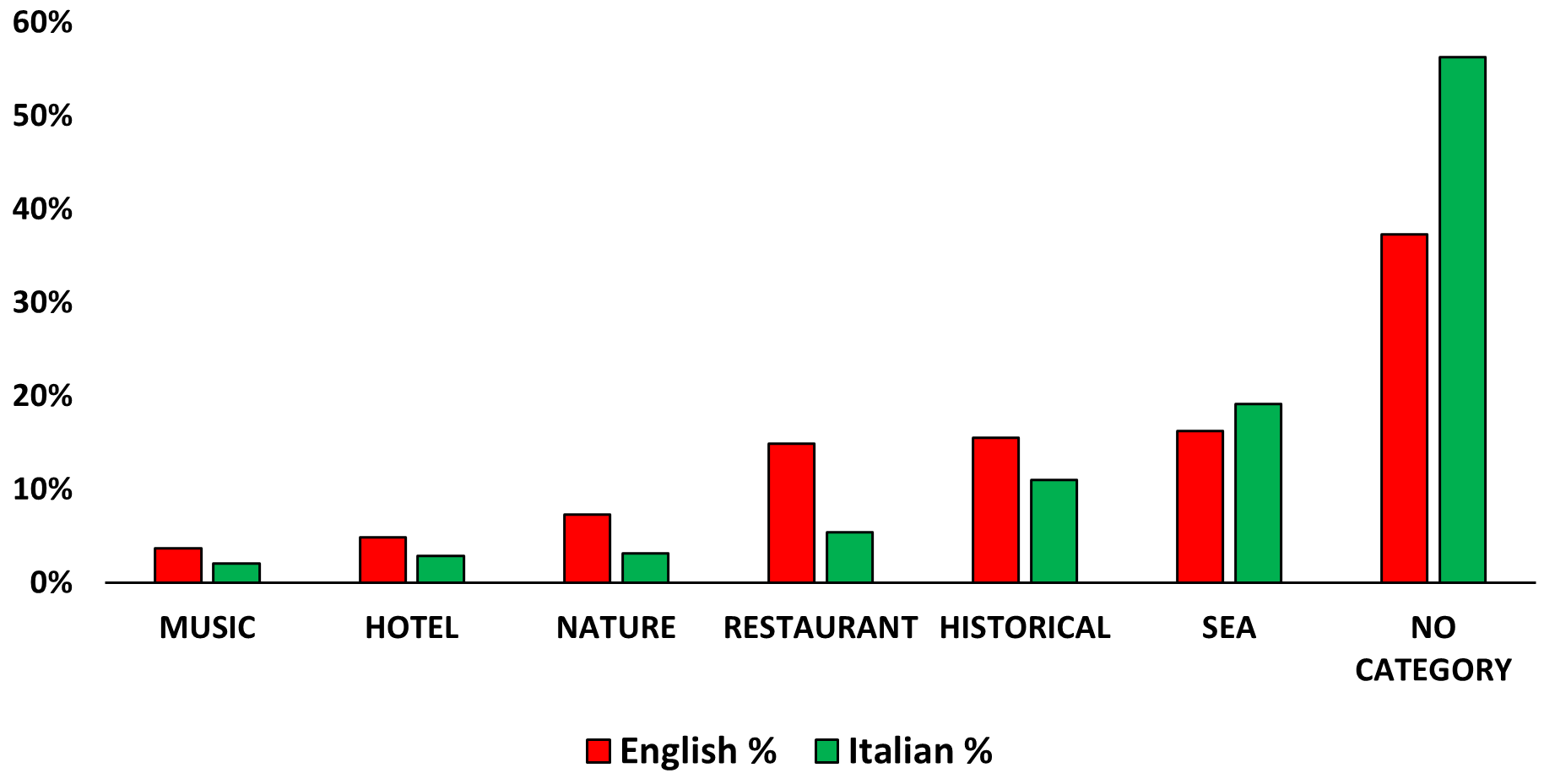}
  \end{center}
  \caption{Category barplot}
  \label{fig: Category barplot for English and Italian}
  \vspace{-10pt}
\end{figure}

In order to explore and gain more insights from the cleaned dataset, we divided it into seven sub categories, \textbf{Sea} (sea and summer holidays) \textbf{Historical} (Ancient remains and monuments), \textbf{Nature} (parks, natural reserves), \textbf{Hotel} (accommodation facilities), \textbf{Restaurant} (restaurants or traditional Puglia's dishes), \textbf{Music} (musical events, concerts or disco parties) and \textbf{General Tourism} (about tourism but not in any specific way) by using keywords and regular expression. Moreover, we used regular expression to find the most often mentioned cities, touristic attractions, hashtags and adjectives.  
The figure~\ref{fig: Category barplot for English and Italian} shows the percentage of each sub category. Mostly of tweets belongs to \textbf{General Tourism} (around 37\% for English and around 56\% for Italian).\\
    
Table ~\ref{tab:The Most frequent words of sub categories} shows the top 15 frequent word of each sub category, sorted in decreasing order by frequency. By observing this table, it is clear that the categorization was precisely correct since all words are related to each category and it also demonstrates what people mostly talk about on Twitter.
For example, the English results show that, for the sea category, there are many tweets that mention the Adriatic coast. The most often cited city is \textit{Polignano a Mare}, a charming town famous throughout the world for his beauty. Similarly, a large part of Italian's tweets speak about \textit{Taranto}, one of the biggest and beautiful city of Puglia situated on Ionian coast. Moreover, by observing the results for the nature category, we can notice that there are many references to the olive trees, iconic symbol of Puglia. For each sub category of tourism, we found the most often mentioned cities, touristic attractions and hashtags.
\begin{table}[ht]
	\centering
	\scalebox{0.7}{
		\begin{tabular}{|p{1.6cm}|p{7.5cm}|p{7.5cm}|}
			\hline
			\textbf{Category} & \textbf{English Keywords} &  \textbf{Italian Keywords} \\
			\hline
			\hline
			Sea &  sea, beach, summer, apulia, travel, coast, adriatic, mare, place, sunset, town, go, holiday, visit, polignano\_mare &  mare, taranto, weareinpugliapuglia, madeintaranto, caraibidipuglia, spiaggia, sea, travel, estate, vacanza, borghitalia, borghipiubelliditalia, otranto, apulia, cittadimare \\
			\hline
			Historical & trulli, alberobello, travel, apulia, town, lecce, southern, visit, one, house, like, city, place, art, italian & alberobello, weareinpugliapuglia, taranto, trulli, travel, trullo, turismo, storia, parte, viaggiare, bellezza, madeintaranto, apulia, arte, lecce\\
			\hline
			Nature & tree, olive, year, nature, travel, think, apulia, lecce, italian, south, place, one, southern, visit, like &  weareinpugliapuglia, natura, nature, travel, taranto, turismo, albero, apulia, otranto, parco, valle, viaggiare, lecce, vacanza, ulivo \\
			\hline
			Hotel & hotel, villa, masseria, apulia, village, travel, pool, holiday, italian, day, resort, one, ostuni, go, stay & hotel, villa, masseria, vacanza, bedandbreakfast, weareinpugliapuglia, gallipoli, apuliatravels, lacasadegliartisti, piscina, villasatyria, villaggio, vacanze, resort \\
			\hline
			Restaurant & wine, food, bari, italian, like, one, travel, go, make, primitivo, restaurant, burrata, visit, pasta, eat &  ristorante, food, danonnadelia, pranzo, taranto, weareinpugliapuglia, menu, foodlovers, gastronomia\_nonna, seafood, seafoodlovers, gastronomia, lunch, tagsforlike, tagsforfollow \\
			\hline
			Music & music, dance, travel, photography, sound, lecce, apulia, summer, year, dj, italian, travelphotography, concert, sunset, time & photography, musica, weareinpugliapuglia, canzone, photo, anno, concerto, apulia, lecce, picture, estate, regionepuglia, taranta, rocchettasantantonio, picoftheday \\
			\hline
			General Tourism & travel, visit, place, south, holiday, go, love, like, apulia, one, gargano, lecce, year, summer, weareinpugliapuglia & taranto, weareinpugliapuglia, madeintaranto, vacanza, turismo, travel, viaggiare, anno, traveling, instatravel, lecce, andare, massafra, picoftheday, estate  \\
			\hline
		\end{tabular}}
	\caption{The Most frequent words of sub categories}
	\label{tab:The Most frequent words of sub categories}
	\vspace{-20pt}
\end{table}

\begin{figure}[!h]
   \begin{minipage}{0.42\textwidth}
     \centering
     \includegraphics[width=1.0\linewidth]{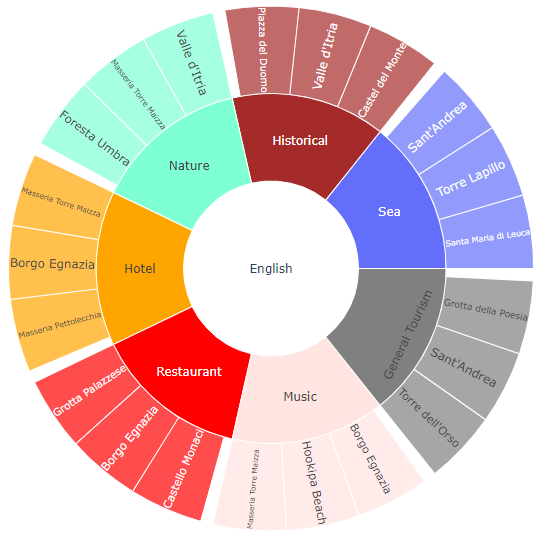}
     \caption{3 Most frequent Tourist Attractions for English}\label{Fig:attr en}
   \end{minipage}\hfill
   \begin{minipage}{0.42\textwidth}
     \centering
     \includegraphics[width=1.0\linewidth]{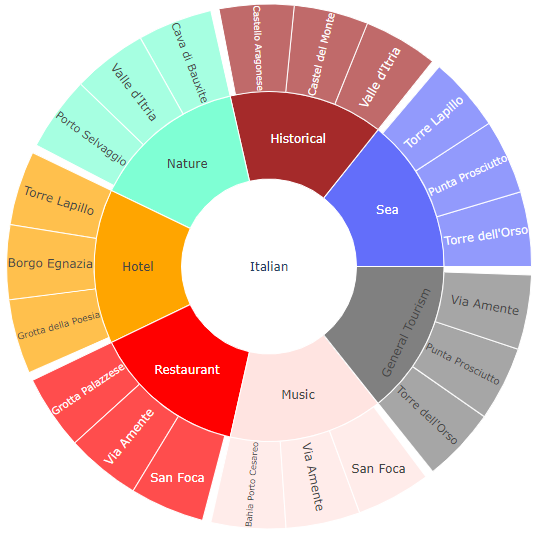}
     \caption{3 Most frequent Tourist Attractions for Italian}\label{Fig:attr it}
   \end{minipage}
\end{figure}

\begin{figure}[!h]
   \begin{minipage}{0.42\textwidth}
   \vspace{-10pt}
     \centering
     \includegraphics[width=1.0\linewidth]{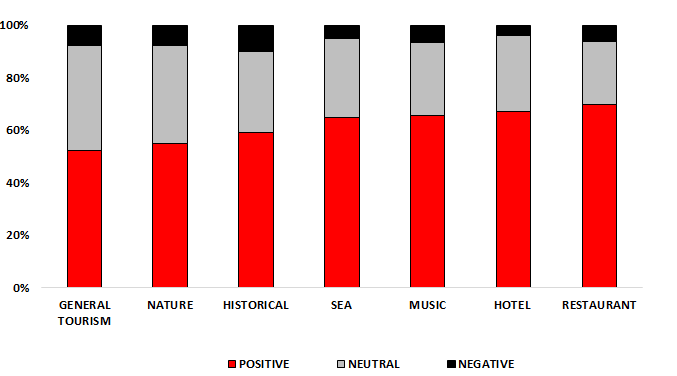}
     \caption{Vader sentiment}\label{Fig:Vader sentiment}
     \vspace{-10pt}
   \end{minipage}\hfill
   \begin{minipage}{0.42\textwidth}
   \vspace{-20pt}
     \centering
     \includegraphics[width=1.0\linewidth]{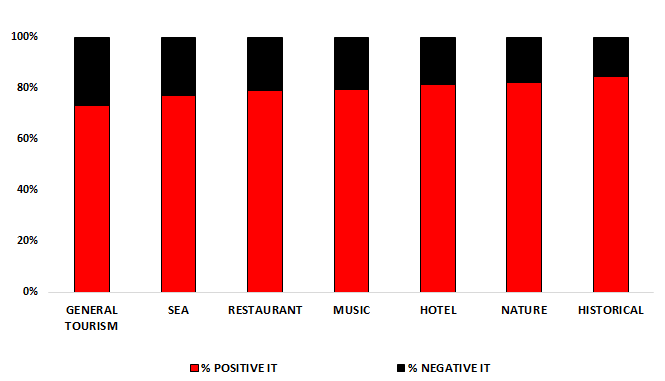}
     \caption{UmBERTo sentiment analysis}\label{Fig: Umberto sentiment}
     \vspace{-10pt}
   \end{minipage}
\end{figure}

In the figures ~\ref{Fig:attr en} and ~\ref{Fig:attr it} we only show the most frequent tourist attractions considering both English and Italian dataset. For example we can note that the town \textit{Santa Maria di Leuca} is the most frequently cited in the English dataset for the sea category, while \textit{Torre dell'Orso} is the most often mentioned one in the Italian dataset. Both are really famous sea places for summer holidays in Puglia.\\
The last analysis presented here is Sentiment Analysis split for category. For the English dataset we used VADER Sentiment\footnote{https://github.com/cjhutto/vaderSentiment} to find the tweets score (figure ~\ref{Fig:Vader sentiment}), while for the Italian dataset we applied UmBERTo \footnote{https://github.com/musixmatchresearch/umberto} (figure ~\ref{Fig: Umberto sentiment}). In the first case the model returns a number between -1 and 1 associated to each tweet  (negative from -1 to 0 and positive from 0 to 1), while for UmBERTo the output associated to each text is a string \textit{positive} or \textit{negative}. \\
We can observed that the sentiment for each category is always positive (for both languages). It means that people usually use Twitter to share their experience about places when it is positive, in contrast with other platforms like TripAdvisor or Google.\\
\\

\section{Social network analysis (SNA) }
\label{sec:Social network analysis }
Social network analysis (SNA)~\cite{Tabassum2018} is a set of techniques used to investigate social structures through the use of networks and graphs made of nodes and ties, edges or links that connect them. Examples of social structures commonly visualized through SNA include social media networks, memes spread, information circulation, disease transmission, etc. In our analysis we used SNA to model the twitter information flow about tourism in Puglia and to do so we used different measures and algorithms such as: \footnote{https://cambridge-intelligence.com/keylines-faqs-social-network-analysis/}
\begin{itemize}
  \item \textbf{Betweenness Centrality}:  Betweenness centrality measures the number of times a node lies on the shortest path between other nodes.
  \item \textbf{Degree Centrality}: Degree centrality assigns an importance score based simply on the number of links held by each node.
  \item \textbf{Eigenvector Centrality}: Like degree centrality, EigenCentrality measures a node’s influence based on the number of links it has to other nodes in the network. EigenCentrality then goes a step further by also taking into account how well connected a node is, and how many links their connections have, and so on through the network.
  \item \textbf{Closeness Centrality}: Closeness centrality scores each node based on their ‘closeness’ to all other nodes in the network.
  \item \textbf{Label Propagation}:
  In the label propagation-based algorithms, a label is propagated to various nodes existing in the network. Each node gains the label possessed by a maximum number of its neighboring nodes.
  \item \textbf{Greedy Modularity}: Modularity is a measure of the structure of networks or graphs which measures the strength of division of a network into modules. Networks with high modularity have dense connections between the nodes within modules but sparse connections between nodes in different modules. 
\end{itemize}
\subsection{Connections between cities and touristic attractions}
Since our aim was to inspect how the information about tourism in Puglia flow on Twitter, we needed to extract from tweets both the main cities and touristic attractions of Puglia, and the opinion of Twitter users about them expressed through tweets. On one hand we inspected the connections between towns and tourist destinations in order to understand how often they are mentioned together. On the other hand we tracked what people claimed about those places, by analyzing both the adjective and the hashtags associated to them.  

\begin{figure}[h]
\includegraphics[width=17cm]{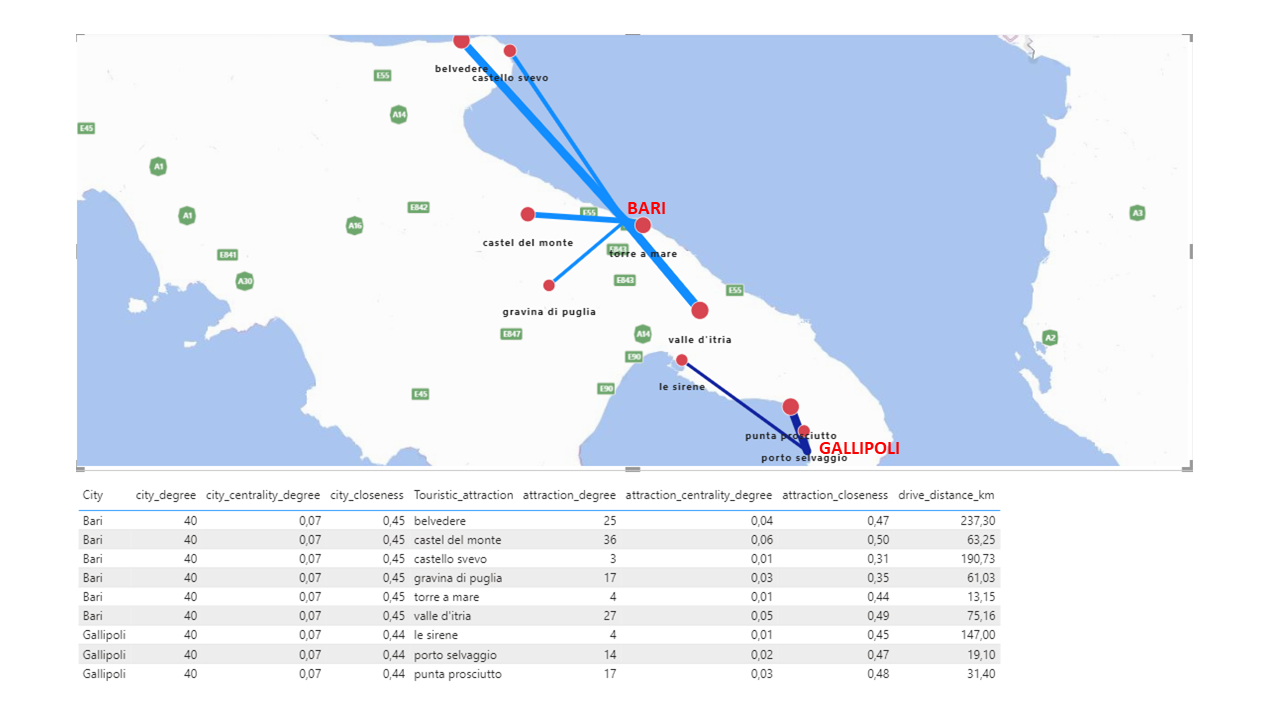}
\centering
\caption{Connection between cities and touristic attractions}
\label{fig:Connection between cities and touristic attractions}
\end{figure}

\begin{figure}[h]
\includegraphics[width=15cm]{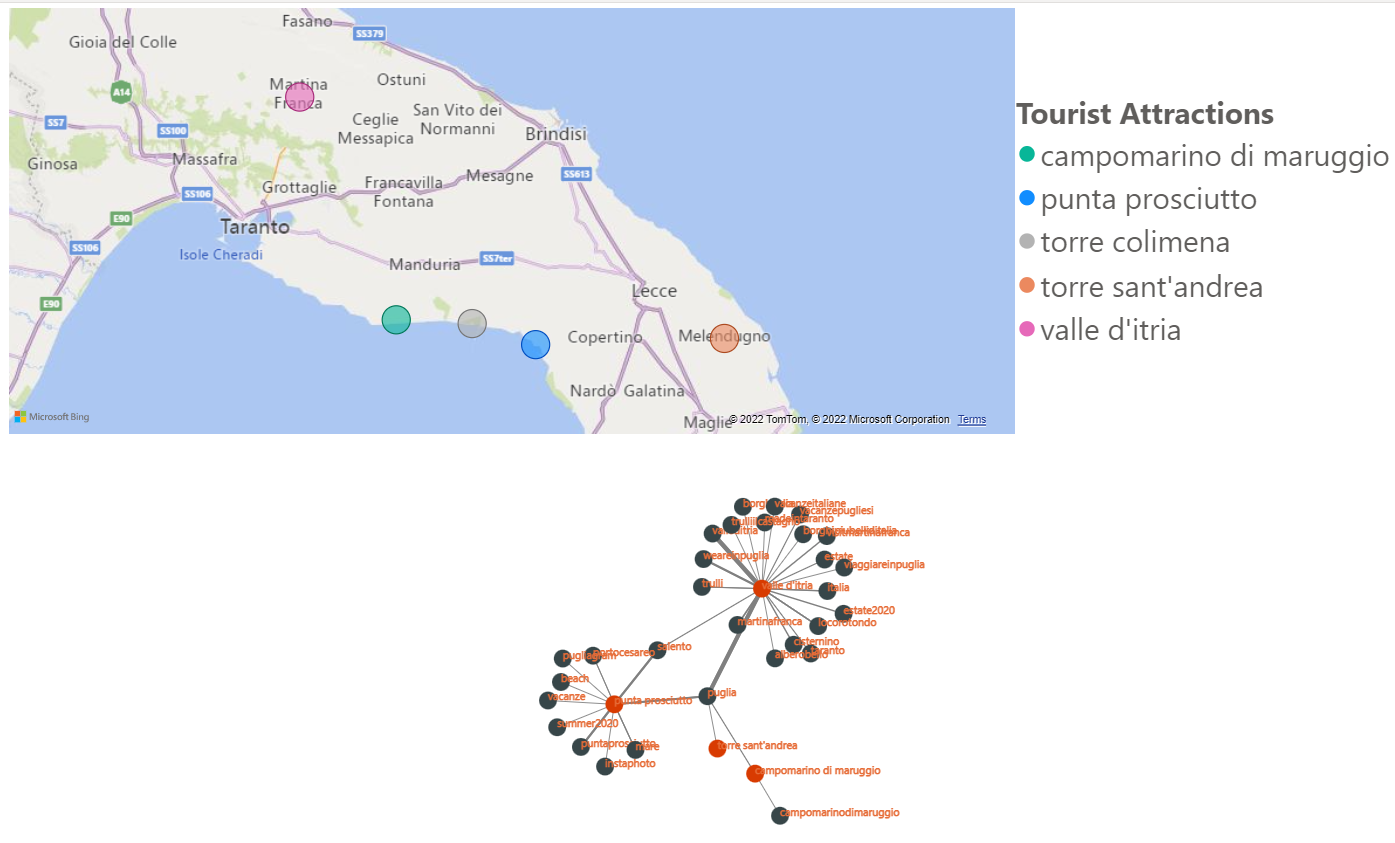}
\centering
\caption{Adjectives and hashtag associated to touristic attractions}
\label{fig:Adjectives and hashtag associated to touristic attractions}
\end{figure}

Figure~\ref{fig:Connection between cities and touristic attractions} is an example of the resulted graph considering the Italian dataset on Puglia maps, where nodes represent cities names and touristic attractions, and links are the connections between them. The bigger the node is, the more often the word associated to it has been mentioned, and the thicker the link is the more often the touristic attraction and the town appear in the same tweets. 
We measured three indexes that represent three different aspects of each node in the graph: \textit{degree}, \textit{centrality degree} and \textit{closeness}.
For a clear demonstration, in the image ~\ref{fig:Connection between cities and touristic attractions} we represents only connections for \textit{Bari}\footnote{https://en.wikipedia.org/wiki/Bari} (violet links) and \textit{Gallipoli}\footnote{https://en.wikipedia.org/wiki/Gallipoli} (orange links), instead the indexes were calculated for the entire graph. 
It is possible to observe that \textit{Bari} and \textit{Gallipoli} have the same numbers of touristic places connected to them (degree = 40). As regards the tourist attractions, we can notice that \textit{Castello Svevo} has only three cities that are connected to it, causing a low closness index (0,31).\\
Analysing the connections between places on Twitter helps us to grasp the more visited places in Puglia, and moreover to inspect which places are visited together. %
Moreover, we tracked tourists opinion about the places they talk about, therefore we analyzed what people say about them, calculating the percentage of positive and negative adjectives. As well as the adjectives, also the hashtags can reveal the general sentiment of a tweet and the opinion of the users. \\
In  figure~\ref{fig:Adjectives and hashtag associated to touristic attractions}, we show an example for \textit{Torre Colimena}, \textit{Valle d'Itria}, \textit{Ponte Girevole}, \textit{Torre Sant'Andrea}, \textit{Campomarino di Maruggio}. For each of them we found the hashtag connected to these touristic destinations, that allow us to inspect which is the topic related to them.

\begin{figure}[h]
\includegraphics[width=8cm, height=4cm]{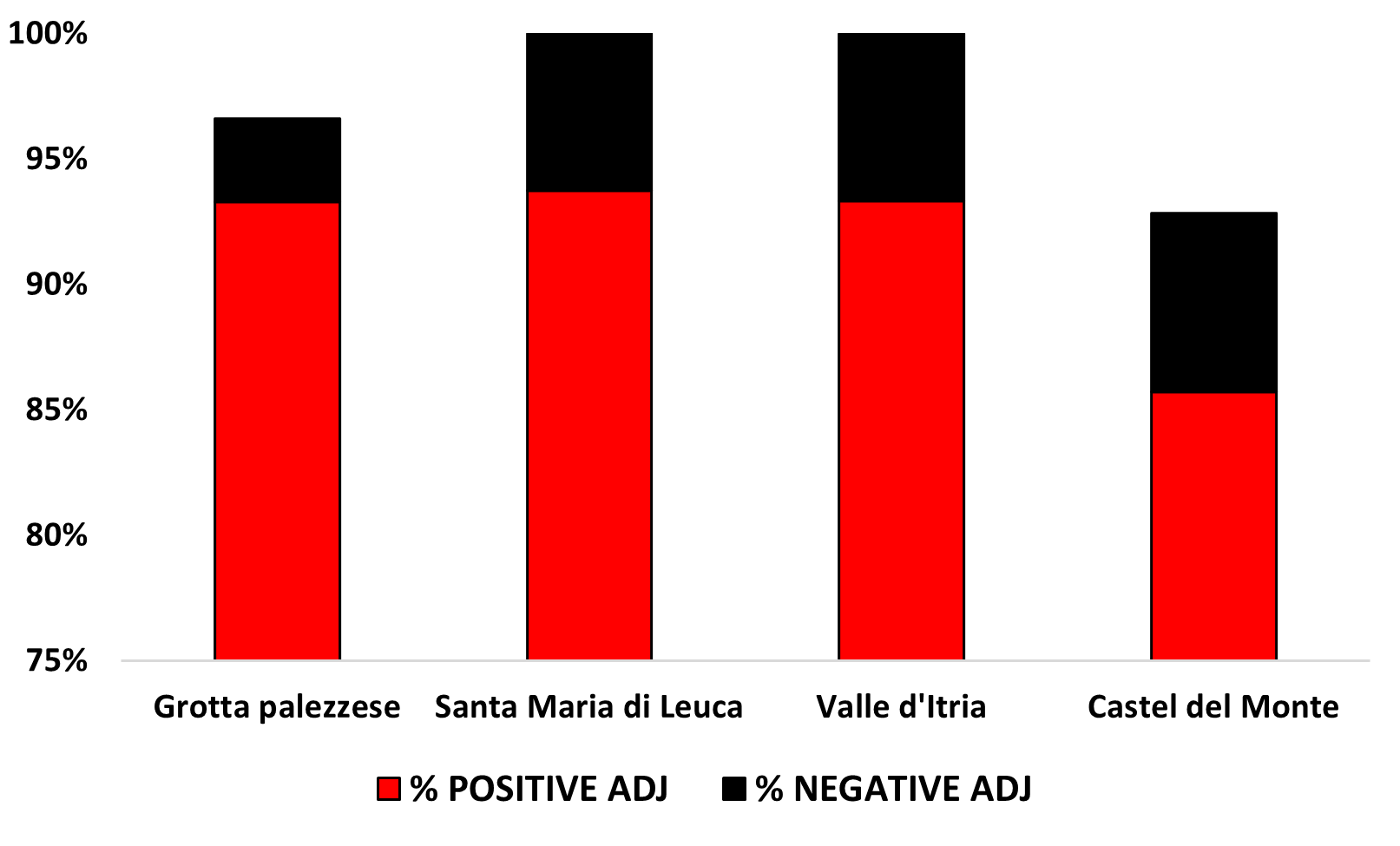}
\centering
\caption{Adjectives associated to touristic attractions}
\label{fig:Adjectives associated to touristic attractions}
\end{figure}

To achieve a qualitative measure that describes the tourists opinion, first we calculated the frequency of both positive and negative adjectives that appear in the same tweets where the touristic attractions are cited, and then we computed the percentage. By observing the figure~\ref{fig:Adjectives associated to touristic attractions} we can notice that the opinions expressed on Twitter are usually positive, as we have already noted in previous analysis.

\subsection{Nodes Importance}

We also analysed our network from another point of view which is exploring words that are carrying the main information through the computation of relevant network metrics. 
There are a series of metrics aimed to measure important nodes, according to different criteria~\cite{pub.1111566433}. Moreover these nodes will always form some kind of groups and clusters in a network. A community is a subset of nodes that interact with each other more than the nodes outside that particular community ~\cite{wasserman1994social}. The nodes within a community are more associated, with more links between themselves and fewer edges with other nodes outside. So here we tried to find out the important nodes in four different network (1:Italian positive, 2:Italian negative, 3:English positive, 4:English negative) by matching the results of centralities and community detection. In other words, not only we want to find important nodes by using centrality and community detection algorithms, but also discover any relation between node centralities and the community detection hub-dominant nodes. With regards of our new analysis we needed to create a new graph which can connect all words together, away from the type of words.

\begin{figure}[H]
\includegraphics[width=5cm, height=4cm]{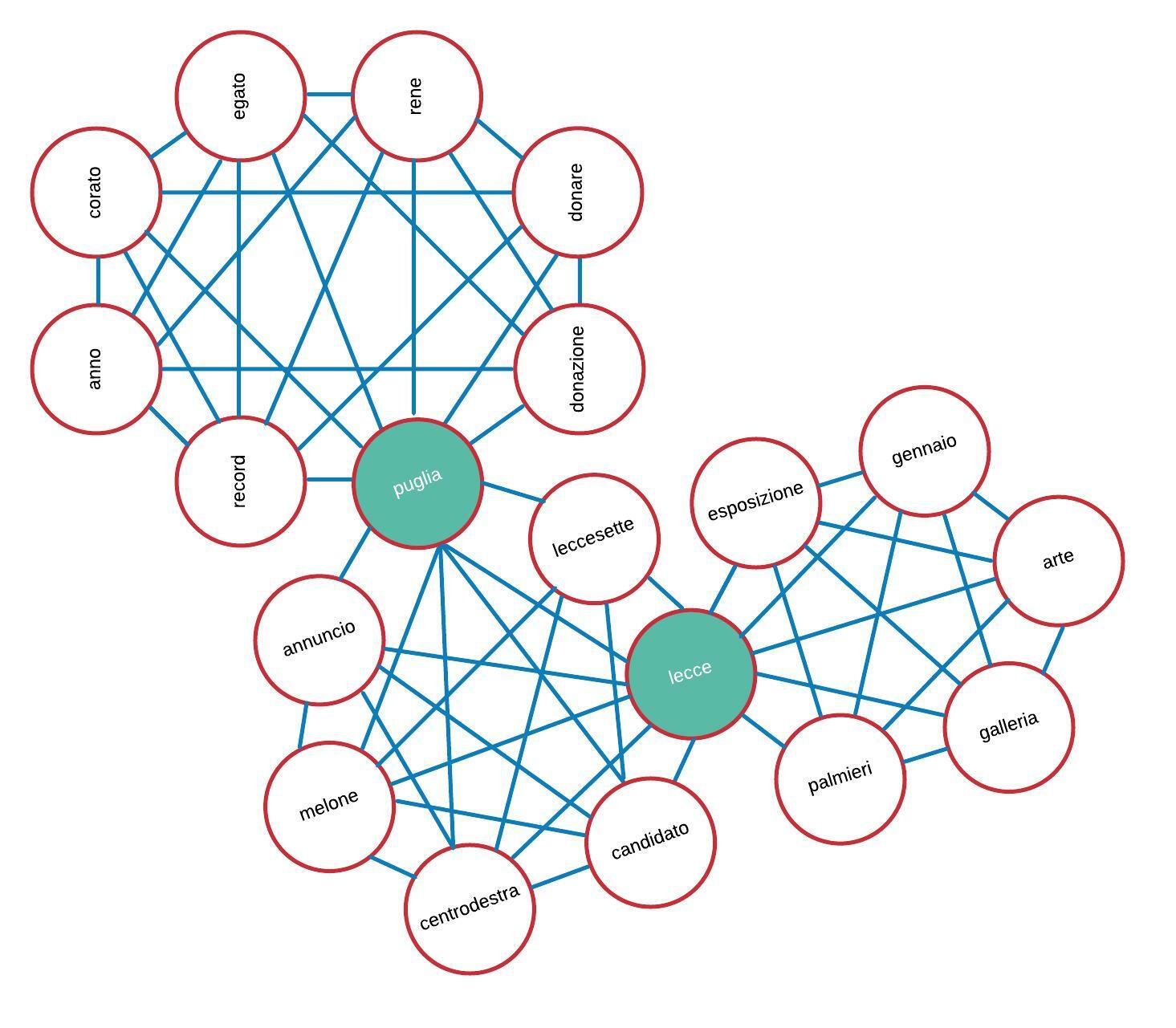}
\centering
\caption{Mesh network}
\label{fig: Mesh network}
\end{figure}


In order to create a new network, we need to identify word pairs. Usually, word pairs refer to pairs of consecutive words found in the text, named n-grams. Here we consider two words as a pair if they appear in a tweet. In this case, we expand our analysis further than ngram and we are able to have a wider knowledge about words and pairs of words that are more used in Twitter about Puglia. 
In this Network, each word is represented by a node and edges represent the existence of word pairs that we mentioned above. The edge's weight and the node's frequency represent the frequency with which these word pairs and nodes are found in the tweets, resulting in a four distinct word pairs network and we assign them as properties for nodes and edges in each distinct network.
Figure~\ref{fig: Mesh network} is an example of how three tweets from the Italian positive dataset will look like when they transfer to a network. By converting tweets into a network, we have a Mesh network for each tweet. A Mesh network or simply meshnet is a network topology in which the infrastructure nodes connect directly, dynamically, and non-hierarchically to as many other nodes as possible.\\ 
Table~\ref{tab:Macroscopic analysis of all networks} shows the characteristics of each network. In terms of nodes and links, the four networks are different in terms of volume. All networks on the positive side have 3.015–3.353 nodes so they are big networks. On the negative side we have 170-657 nodes which shows us we have much smaller networks. Also there is density which is the number of connections a word has divided by the total possible connections a word could have in the network. The density is 0 for a graph without edges and 1 for a complete graph. It ranges from 0.023 to 0.032 on the positive side and 0.035 to 0.084 on the negative sides, and both networks are considered a less complete graph (normal graph). Also we have max and average degree for each network and surprisingly all four network share a close range of average degree.
\begin{table}[ht]
	\centering
	\vspace{-1pt}
	\scalebox{0.7}{
		\begin{tabular}{ | c | c | c | c | c | c |}
			\hline
			{Network} &  {Nodes} & {Edges} & {Density} & {Max Degree} & {Avg Degree}\\
			\hline
			Italian Positive & 3015 & 505 & 0.023 & 190 & 2,65 \\
			\hline
			Italian Negative & 170 & 64 & 0.084 & 37 & 1,72\\
			\hline
			English Positive & 3353 & 452 & 0.032 & 194 & 2,31 \\
			\hline
			English Negative & 657 & 194 & 0.035 & 80 & 2,4\\
			\hline
		\end{tabular}}
	\caption{Macroscopic analysis of all networks}
	\label{tab:Macroscopic analysis of all networks}
\end{table}
\vspace{-10pt}
\subsection{Centralises Measure}
In this phase, we are going to find important words by calculating through the computation of relevant network metrics, such as betweenness, closeness, degree, and eigenvector centrality. These measures can be used to locate nodes representing semantic resources that have the most advantageous positions compared to other nodes in the network.\\
\begin{table}[h]
    \vspace{-1pt}
	\centering
	\scalebox{0.7}{
		\begin{tabular}{ | c | c | c | c | c | c | c | c |}
			\hline
			\multicolumn{4}{|c|}{Italian} & \multicolumn{4}{|c|}{English}\\
			\hline
			\multicolumn{2}{|c|}{Pos} & \multicolumn{2}{c|}{Neg} & 	\multicolumn{2}{c|}{Pos} & 	\multicolumn{2}{c|}{Neg}\\
			\hline
			{Words} &  {Score} & {Words} & {Score} & {Words} &  {Score} & {Words} &  {Score}\\
			\hline
			mare & 0.24 & mare & 0.55 & salento & 0.18 & salento & 0.50 \\
			\hline
			weareinpuglia & 0.19 & vacanza & 0.28 & region & 0.12 & apulia & 0.14\\
			\hline
			italia & 0.16 & italia & 0.24 & travel & 0.12 & region & 0.11\\
			\hline
			italy & 0.15 & italy & 0.14 & beautiful & 0.09 & travel & 0.08\\
			\hline
			taranto & 0.13 & andare & 0.04 & wine & 0.18 & go & 0.07 \\
			\hline
			travel & 0.04 & weareinpuglia & 0.04 & sea & 0.12 & monopoli & 0.06\\
			\hline
			ristorante & 0.03 & taranto & 0.04 & visit & 0.12 & sea & 0.06\\
			\hline
			lecce & 0.03 & anno & 0.03 & love & 0.09 & italian & 0.05\\
			\hline
			vacanza & 0.02 & estate & 0.03 & like & 0.12 & lecce & 0.05\\
			\hline
			apulia & 0.02 & turismo & 0.01 & apulia & 0.09 & southern & 0.05\\
			\hline
	\end{tabular}}
	\caption{Betweenness Centrality}
	\label{tab:Betweenness Centrality}
	\vspace{-10pt}
\end{table}
\textbf{Betweenness:} These nodes are the 'bridges' between other nodes in a network. In other words it’s the gate of entrance to each cluster of nodes. Table~\ref{tab:Betweenness Centrality} shows the top 10 words with the highest overall betweenness centrality. You can guess approximately what clusters we may have, some of them are pretty much clear and for others it is more general
\begin{table}[h]
    \vspace{-1pt}
	\centering
	\scalebox{0.7}{
		\begin{tabular}{ | c | c | c | c | c | c | c | c |}
			\hline
			\multicolumn{4}{|c|}{Italian} & \multicolumn{4}{|c|}{English}\\
			\hline
			\multicolumn{2}{|c|}{Pos} & \multicolumn{2}{c|}{Neg} & 	\multicolumn{2}{c|}{Pos} & 	\multicolumn{2}{c|}{Neg}\\
			\hline
			{Words} & {Score} & {Words} & {Score} & {Words} & {Score} & {Words} & {Score}\\
			\hline
			mare & 0.58 & mare & 0.69 & salento & 0.61 & salento & 0.56 \\
			\hline
			italia & 0.58 & vacanza & 0.58 & travel & 0.59 & travel & 0.47\\
			\hline
			italy & 0.57 & italia & 0.56 & beautiful & 0.58 & apulia & 0.47\\
			\hline
			weareinpuglia & 0.56 & italy & 0.50 & region & 0.58 & region & 0.46\\
			\hline
			travel & 0.52 & sapere & 0.49 & love & 0.55 & italian & 0.44 \\
			\hline
			taranto & 0.52 & andare & 0.49 & visit & 0.55 & southern & 0.44\\
			\hline
			apulia & 0.49 & dire & 0.49 & one & 0.55 & italia & 0.42\\
			\hline
			sea & 0.48 & estate & 0.49 & apulia & 0.55 & lecce & 0.42\\
			\hline
			summer & 0.48 & weareinpuglia & 0.48 & like & 0.54 & town & 0.42\\
			\hline
			photooftheday & 0.48 & vedere & 0.48 & place & 0.54 & summer & 0.42\\
			\hline
		\end{tabular}}
	\caption{Closeness Centrality}
	\label{tab:Closeness Centrality}
	\vspace{-10pt}
\end{table}

\textbf{Closeness:} These words are in favorable positions in the networks to acquire and control vital information and spread information in an efficient manner. You can see in table~\ref{tab:Closeness Centrality} that in the Italian network on both side, positive and negative, the word ("mare":sea) is the more central word, thus it’s closer to all other words. Also the word \textit{Salento} appeared more in the center in the English network. These words can give us the idea of what people more focus on, while they tweet about tourism in Italian

\begin{table}[h]
    \vspace{-1pt}
	\centering
	\scalebox{0.7}{
		\begin{tabular}{ | c | c | c | c | c | c | c | c |}
			\hline
			\multicolumn{4}{|c|}{Italian} & \multicolumn{4}{|c|}{English}\\
			\hline
			\multicolumn{2}{|c|}{Pos} & 	\multicolumn{2}{c|}{Neg} & 	\multicolumn{2}{c|}{Pos} & 	\multicolumn{2}{c|}{Neg}\\
			\hline
			{Words} & {Score} & {Words} & {Score} & {Words} &  {Score} & {Words} &  {Score}\\
			\hline
			mare & 0.37 & mare & 0.58 & salento & 0.43 & salento & 0.41 \\
			\hline
			italy & 0.35 & vacanza & 0.41 & travel & 0.19 & travel & 0.19\\
			\hline
			italia & 0.34 & andare & 0.26 & beautiful & 0.19 & region & 0.19\\
			\hline
			weareinpuglia & 0.34 & italia & 0.26 & region & 0.18 & apulia & 0.18\\
			\hline
			taranto & 0.28 & taranto & 0.25 & visit & 0.15 & go & 0.15 \\
			\hline
			travel & 0.24 & anno & 0.22 & love & 0.12 & italian & 0.12\\
			\hline
			apulia & 0.13 & italy & 0.17 & apulia & 0.12 & italia & 0.12\\
			\hline
			sea & 0.12 & sapere & 0.15 & one & 0.12 & sea & 0.12\\
			\hline
			summer & 0.12 & turismo & 0.15 & like & 0.11 & bari & 0.11\\
			\hline
			photooftheday & 0.12 & viaggiare & 0.14 & place & 0.10 & summer & 0.10\\
			\hline
		\end{tabular}}
	\caption{Degree Centrality}
	\label{tab:Degree Centrality}
	\vspace{-10pt}
\end{table}
\textbf{Degree:} Degree centrality is the simplest measure of node connectivity. It is a good measure for finding popular words, words who are likely to hold most information or it can quickly connect with the wider network. As you can see in table~\ref{tab:Degree Centrality} the result are very similar to what we have in closness centrality. Words ("mare":sea) and \textit{Salento} are the most vital nodes in our networks.

\begin{table}[H]  
    \vspace{-1pt}
	\centering
	\scalebox{0.7}{
		\begin{tabular}{ | c | c | c | c | c | c | c | c |}
			\hline
			\multicolumn{4}{|c|}{Italian} & \multicolumn{4}{|c|}{English}\\
			\hline
			\multicolumn{2}{|c|}{Pos} & 	\multicolumn{2}{c|}{Neg} & 	\multicolumn{2}{c|}{Pos} & 	\multicolumn{2}{c|}{Neg}\\
			\hline
			{Words} & {Score} & {Words} & {Score} & {Words} & {Score} & {Words} & {Score}\\
			\hline
			italy & 0.26 & mare & 0.42 & beautiful & 0.18 & salento & 0.36 \\
			\hline
			italia & 0.26 & vacanza & 0.31 & travel & 0.18 & travel & 0.25\\
			\hline
			mare & 0.24 & andare & 0.27 & salento & 0.18 & apulia & 0.24\\
			\hline
			travel & 0.22 & italia & 0.24 & region & 0.17 & region & 0.23\\
			\hline
			weareinpuglia & 0.22 & anno & 0.24 & visit & 0.16 & italian & 0.22 \\
			\hline
			taranto & 0.16 & sapere & 0.21 & love & 0.16 & go & 0.18\\
			\hline
			sea & 0.15 & taranto & 0.18 & one & 0.16 & italia & 0.17\\
			\hline
			summer & 0.15 & dire & 0.18 & place & 0.16 & bari & 0.17\\
			\hline
			photooftheday & 0.15 & italy & 0.18 & apulia & 0.16 & summer & 0.16\\
			\hline
			instagood & 0.14 & vedere & 0.16 & like & 0.16 & southern & 0.15\\
			\hline
		\end{tabular}}
	\caption{Eigenvector Centrality}
	\label{tab:Eigenvector Centrality}
	\vspace{-10pt}
\end{table}
\textbf{Eigenvector:} EigenCentrality can identify nodes with influence over the whole network, not just those directly connected to it. It is a good SNA score which is useful to understanding a human social networks like twitter. Table~\ref{tab:Eigenvector Centrality} tell us that in the Italian network the word ("mare":sea) has the highest negative influence on our network, while the word \textit{Salento} has the same role in the English network.\\ 

Now we are going to analyse the community detection. As we said before we wanted to see if there is any relation between those important nodes according to centralities measures and hub-dominant nodes of communities detection.
\subsection{Community Detection}
Community detection, also called graph partition, helps us to reveal the hidden relations among the nodes in the network. The selection of the community detection tool as well as the interpretation of its results is strongly dependent on both the user’s goals. Each community algorithm potentially solves a slightly different version of the problem.\\  
Greedy modularity and Label propagation are methods that we used here. Both algorithms split networks into several communities which some of them only have one or two nodes that basically makes them an outlier. So as to achieve communities that contain a big number of nodes, we calculated the standard deviation of communities in each network and put it as the threshold here. So base on that, we only choose communities that have number of nodes more than the threshold. Table~\ref{tab:Number of Communities Detection} shows the number of communities for each network and the number of community that we have considered.
\begin{table}[H]
	\centering
	\scalebox{0.8}{
		\begin{tabular}{ | c | c | c | c | c | }
			\hline
			{Network} & {Greedy modularity} & {Chosen Community} & {Label propagation} & {Chosen Community}\\
			\hline
			Italian Positive & 19 & 2 & 21 & 1\\
			\hline
			Italian Negative & 3 & 1 & 9 & 3\\
			\hline
			English Positive & 10 & 4 & 17 & 1\\
			\hline
			English Negative & 16 & 3 & 17 & 1\\
			\hline
		\end{tabular}}
	\caption{Number of Communities Detection}
	\label{tab:Number of Communities Detection}
	\vspace{-10pt}
\end{table}
The quality of community detection in both algorithms are high enough, however as you can see in figures below, the Greedy modularity worked slightly better in terms of splitting networks into groups (Due to the size of our networks, here we only shows the negative part of our network).\\
\\
\begin{figure}[ht]
   \begin{minipage}{0.48\textwidth}
     \vspace{-10pt}
     \centering
     \includegraphics[width=.7\linewidth]{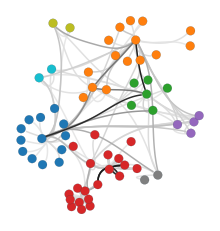}
     \caption{label Propagation-Italian Negative}\label{Fig:label propagation - Italian Negative}
     \vspace{-10pt}
   \end{minipage}\hfill
   \begin{minipage}{0.48\textwidth}
     \vspace{-10pt}
     \centering
     \includegraphics[width=.7\linewidth]{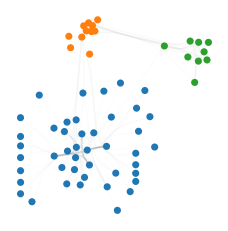}
     \caption{Greedy Modularity-Italian Negative}\label{Fig:Greedy Modularity - Italian Negative}
     \vspace{-10pt}
   \end{minipage}
\end{figure}
\begin{figure}[H]
   \begin{minipage}{0.48\textwidth}
     \vspace{-1pt}
     \centering
     \includegraphics[width=.7\linewidth]{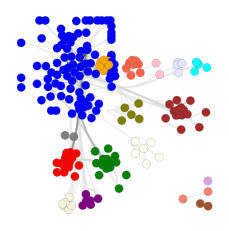}
     \caption{label propagation-English Negative}\label{Fig:label propagation - English Negative}
     \vspace{-10pt}
   \end{minipage}\hfill
   \begin{minipage}{0.48\textwidth}
     \vspace{-20pt}
     \centering
     \includegraphics[width=.7\linewidth]{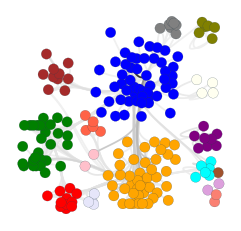}
     \caption{Greedy Modularity-English Negative}\label{Fig:Greedy Modularity - English Negative}
     \vspace{-10pt}
   \end{minipage}
\end{figure}
\begin{wraptable}{r}{0.5\textwidth}
    \vspace{-20pt}
	\centering
	\scalebox{0.5}{
		\begin{tabular}{ | c | c | c |}
			\hline
			{Network} & {Communities name} & {Hub-dominant}\\
			\hline
			Italian Positive & Group-1 & mare\\
			\hline
			Italian Positive & Group-2 & italia\\
			\hline
			Italian Negative & Group-1 & mare\\
			\hline
			English Positive & Group-1 & salento\\
			\hline
			English Positive & Group-2 & travel\\
			\hline
			English Positive & Group-3 & region\\
			\hline
			English Positive & Group-4 & apulia\\
			\hline
			English Negative & Group-1 & salento\\
			\hline
			English Negative & Group-2 & travel\\
			\hline
			English Negative & Group-3 & apulia\\
			\hline
		\end{tabular}}
	\caption{Hub-dominant nodes}
	\label{tab:Hub-dominant nodes}
	\vspace{-10pt}
\end{wraptable}
To compare our results in community detection with centralities measure we need to find the hub-dominant nodes in each community. There is no specific definition of how we should find the hub-dominant nodes in communities, but in most researches, the node with highest degree centrality in each community will be consider as the hub-dominant node. In our case we calculated the degree centrality in each chosen community that we got from Greedy modularity. As you can see in table~\ref{tab:Hub-dominant nodes}, almost all hub-dominant nodes also have the highest centrality measures in our previous analysis. It proved our hypothesis, that each important node will likely gather other nodes around itself and makes group of nodes which each group have its own context. By investigating more on the these context, we may able to use the community detection to categorise our dataset from the beginning.

\section{Conclusion and future works}
\label{sec:conclusion}
\begin{figure}[H]
   \begin{minipage}{0.48\textwidth}
    \vspace{-1pt}
     \centering
     \includegraphics[width=5cm, height=4cm]{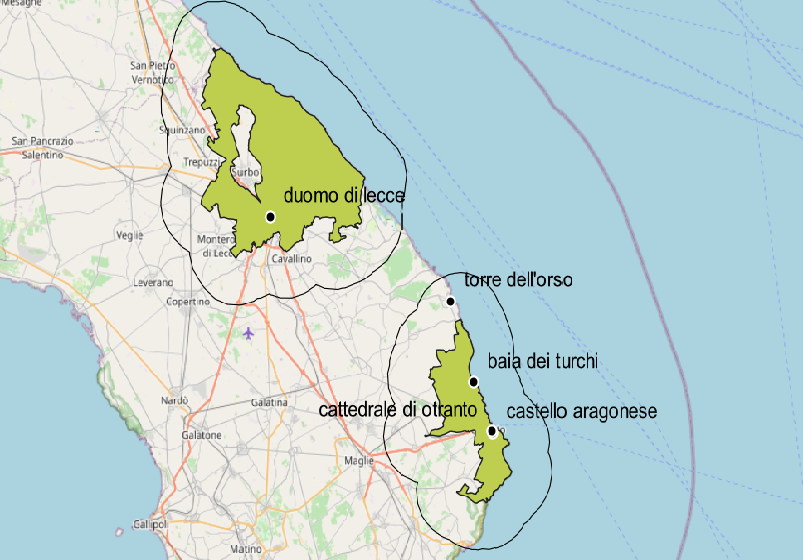}
     \caption{QGis Map-connections between cities and touristic attractions}
     \label{Fig:QGis Map-connections between cities and touristic attractions}
     \vspace{-10pt}
   \end{minipage}\hfill
   \begin{minipage}{0.48\textwidth}
   \vspace{-20pt}
     \centering
     \includegraphics[width=5cm, height=4cm]{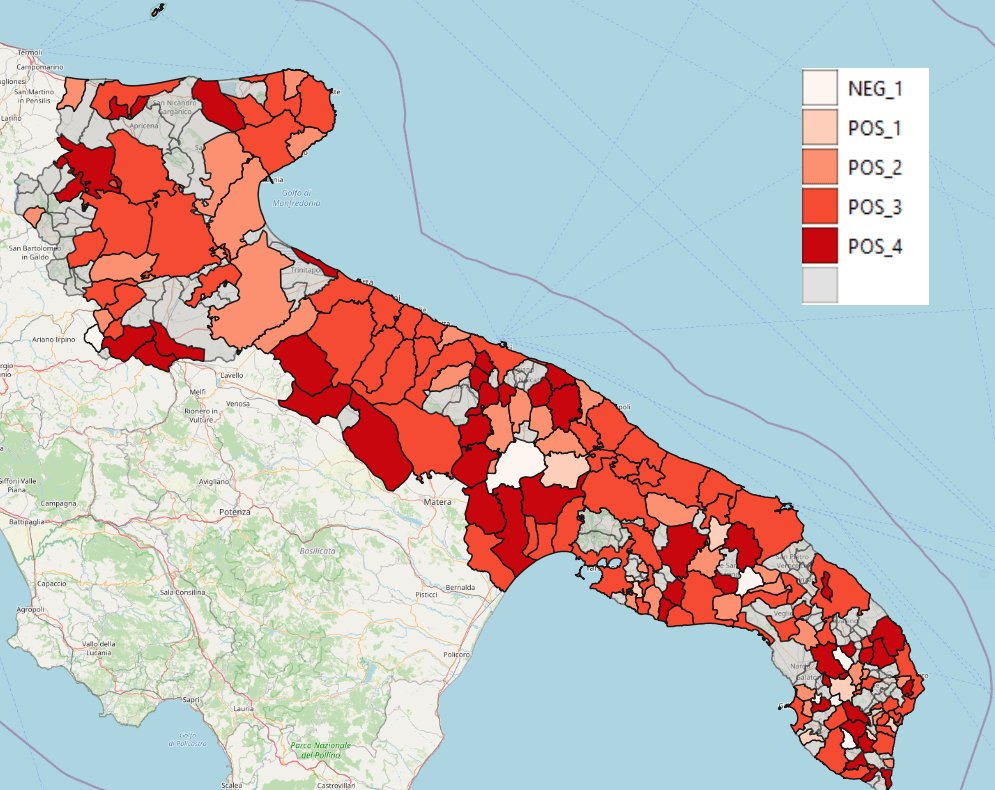}
     \caption{QGis Map-Vader Sentiment}
     \label{Fig:QGis Map-Vader Sentiment}
     \vspace{-10pt}
   \end{minipage}
\end{figure}

The goal of this work was to find how information about tourism in Puglia flow, thorough Twitter, by using Social Network Analysis. In order to achieve this aim, first we had to extract data related to domain we were interested in. So we proposed an approach to filter out only data about tourism in Puglia, by running a semi-automatic method that yet requires a human intervention.\\
The results of cleaning phase were quite convincing, since in all analysis that we did after we have found only words related to tourism topic, despite the fact that the original dataset covered different arguments such as politics, Covid-19 and so on.\\
From the cleaned dataset, we extracted some interesting information about tourism, by splitting tweets in several categories. This exploration phase helped us to understand how to exploit the details about the cities and touristic attractions of Puglia, by transforming tweets into network.
This analysis helped to inspect the correlation between tourist sites. Moreover the links between places in the network that are geographically near would act like a recommendation engine, which can suggest to tourists a new destination to see (Fig.\ref{Fig:QGis Map-connections between cities and touristic attractions}), and the sentiment analysis that we did on the network can also guide tourist to evaluate the destination where they intend to go (Fig.\ref{Fig:QGis Map-Vader Sentiment}).\\
In addition, finding important nodes helped us to detect how the information flow in our network and we were able to see which words have been used to describe tourism situation in Puglia. By following this approach over time we can find out how people's opinion would change about it.\\
In order to simplify the entire workflow of cleaning we are pursuing different ways. On one hand we are exploiting the labeled dataset (touristic and no touristic tweets) to train a supervised classifier (8) that helps us to label the tweets in a fully automatic way. On the other hand, we are going to use community detection from the beginning to categorize tweets in different clusters according to their context and analysing their evolution over the time.
Moreover, in the future we are going to expand this analysis throughout networks, by considering all categories that we have found here, and by analyzing each category's evolution over the time.\\ 

\section*{Acknowledgments}
\vspace{-6pt}
\textbf{Funding/Support:} This work was supported by the POR PUGLIA FESR 2014-2020 project C-BAS "Customer Behaviour Analysis System".
\bibliographystyle{unsrt}
\bibliography{references.bib}

\begin{thebibliography}{10}

\bibitem{meyer2007understanding}
Christopher Meyer, Andre Schwager, et~al.
\newblock Understanding customer experience.
\newblock {\em Harvard business review}, 85(2):116, 2007.

\bibitem{kumar2014twitter}
Shamanth Kumar, Fred Morstatter, and Huan Liu.
\newblock {\em Twitter data analytics}.
\newblock Springer.

\bibitem{antonakaki2021survey}
Despoina Antonakaki, Paraskevi Fragopoulou, and Sotiris Ioannidis.
\newblock A survey of twitter research: Data model, graph structure, sentiment
  analysis and attacks.
\newblock {\em Expert Systems with Applications}, 164:114006, 2021.

\bibitem{khan2021twitter}
Hikmat~Ullah Khan, Shumaila Nasir, Kishwar Nasim, Danial Shabbir, and Ahsan
  Mahmood.
\newblock Twitter trends: a ranking algorithm analysis on real time data.
\newblock {\em Expert Systems with Applications}, 164:113990, 2021.

\bibitem{hassonah2020efficient}
Mohammad~A Hassonah, Rizik Al-Sayyed, Ali Rodan, Al-Zoubi Ala’M, Ibrahim
  Aljarah, and Hossam Faris.
\newblock An efficient hybrid filter and evolutionary wrapper approach for
  sentiment analysis of various topics on twitter.
\newblock {\em Knowledge-Based Systems}, 192:105353, 2020.

\bibitem{yigitcanlar2021smart}
Tan Yigitcanlar, Nayomi Kankanamge, and Karen Vella.
\newblock How are smart city concepts and technologies perceived and utilized?
  a systematic geo-twitter analysis of smart cities in australia.
\newblock {\em Journal of Urban Technology}, 28(1-2):135--154, 2021.

\bibitem{garcia2021topic}
Klaifer Garcia and Lilian Berton.
\newblock Topic detection and sentiment analysis in twitter content related to
  covid-19 from brazil and the usa.
\newblock {\em Applied Soft Computing}, 101:107057, 2021.

\bibitem{singh2016role}
Tajinder Singh and Madhu Kumari.
\newblock Role of text pre-processing in twitter sentiment analysis.
\newblock {\em Procedia Computer Science}, 89:549--554, 2016.

\bibitem{jianqiang2017comparison}
Zhao Jianqiang and Gui Xiaolin.
\newblock Comparison research on text pre-processing methods on twitter
  sentiment analysis.
\newblock {\em IEEE Access}, 5:2870--2879, 2017.

\bibitem{symeonidis2018comparative}
Symeon Symeonidis, Dimitrios Effrosynidis, and Avi Arampatzis.
\newblock A comparative evaluation of pre-processing techniques and their
  interactions for twitter sentiment analysis.
\newblock {\em Expert Systems with Applications}, 110:298--310, 2018.

\bibitem{naseem2020survey}
Usman Naseem, Imran Razzak, and Peter~W Eklund.
\newblock A survey of pre-processing techniques to improve short-text quality:
  a case study on hate speech detection on twitter.
\newblock {\em Multimedia Tools and Applications}, pages 1--28, 2020.

\bibitem{reyes2018understanding}
Ana Reyes-Menendez, Jos{\'e}~Ram{\'o}n Saura, and Cesar Alvarez-Alonso.
\newblock Understanding\# worldenvironmentday user opinions in twitter: A
  topic-based sentiment analysis approach.
\newblock {\em International journal of environmental research and public
  health}, 15(11):2537, 2018.

\bibitem{boon2020public}
Sakun Boon-Itt and Yukolpat Skunkan.
\newblock Public perception of the covid-19 pandemic on twitter: Sentiment
  analysis and topic modeling study.
\newblock {\em JMIR Public Health and Surveillance}, 6(4):e21978, 2020.

\bibitem{ordun2020exploratory}
Catherine Ordun, Sanjay Purushotham, and Edward Raff.
\newblock Exploratory analysis of covid-19 tweets using topic modeling, umap,
  and digraphs.
\newblock {\em arXiv preprint arXiv:2005.03082}, 2020.

\bibitem{yin2020detecting}
Hui Yin, Shuiqiao Yang, and Jianxin Li.
\newblock Detecting topic and sentiment dynamics due to covid-19 pandemic using
  social media.
\newblock In {\em International Conference on Advanced Data Mining and
  Applications}, pages 610--623. Springer, 2020.

\bibitem{blei2003latent}
David~M Blei, Andrew~Y Ng, and Michael~I Jordan.
\newblock Latent dirichlet allocation.
\newblock {\em the Journal of machine Learning research}, 3:993--1022, 2003.

\bibitem{shashaj2021cea}
Ariona Shashaj, Davide Stirparo, and Mohammad Kazemi.
\newblock Cea-tm: A customer experience analysis framework based on
  contextual-aware topic modeling approach.
\newblock In {\em IFIP International Conference on Artificial Intelligence
  Applications and Innovations}, pages 659--672. Springer, 2021.

\bibitem{Tabassum2018}
Shazia Tabassum, Fabiola S.~F. Pereira, Sofia Fernandes, and João Gama.
\newblock Social network analysis: An overview.
\newblock {\em WIREs Data Mining and Knowledge Discovery}, 8(5):e1256, 2018.

\bibitem{pub.1111566433}
Yuanzhi Yang, Lei Yu, Zhongliang Zhou, You Chen, and Tian Kou.
\newblock Node importance ranking in complex networks based on multicriteria
  decision making.
\newblock {\em Mathematical Problems in Engineering}, 2019:1--12, 2019.
\newblock https://doi.org/10.1155/2019/9728742.

\bibitem{wasserman1994social}
S.~Wasserman, K.~Faust, M.~Granovetter, Cambridge~University Press,
  D.~Iacobucci, and University of~Cambridge.
\newblock {\em Social Network Analysis: Methods and Applications}.
\newblock Structural Analysis in the Social Sciences. Cambridge University
  Press, 1994.

\end{thebibliography}
\end{document}